# Spin-mechanical coupling in 2D antiferromagnet CrSBr

*Fan Fei[1†], Yulu Mao[2†], Wuzhang Fang[1], Wenhao Liu[5], Jack P. Rollins[1], Aswin L. N. Kondusamy[5], Bing Lv[5], Yuan Ping[1,3,4], Ying Wang[2,1,3], Jun Xiao[1,2,3,*]*

[1] Department of Materials Science and Engineering, University of Wisconsin-Madison, Madison, Wisconsin 53706, USA

[2] Department of Electrical and Computer Engineering, University of Wisconsin-Madison, Madison, Wisconsin 53706, USA

[3] Department of Physics, University of Wisconsin-Madison, Madison, Wisconsin 53706, USA

[4] Department of Chemistry, University of Wisconsin-Madison, Madison, Wisconsin 53706, USA

[5] Department of Physics, University of Texas at Dallas, Richardson, Texas 75080, USA

†Those authors contributed equally.

*Corresponding author(s). E-mail(s): jun.xiao@wisc.edu;

## Abstract

Spin-mechanical coupling is vital in diverse fields including spintronics, sensing and quantum transduction. Two-dimensional (2D) magnetic materials provide a unique platform for investigating spin-mechanical coupling, attributed to their mechanical flexibility and novel spin orderings. However, studying spin-mechanical coupling in 2D magnets presents challenges in probing mechanical deformation and thermodynamic properties change at nanoscale. Here we use nano opto-electro-mechanical interferometry to mechanically detect the phase transition and magnetostriction effect in multilayer CrSBr, an air-stable antiferromagnets with large magnon-exciton coupling. The transitions among antiferromagnetism, spin-canted ferromagnetism and paramagnetism are visualized by optomechanical frequency anomalies. Nontrivial magnetostriction coefficient $2.3 \times 10^{-5}$ and magnetoelastic coupling strength on the order of $10^6$ J/m$^3$ have been found. Moreover, we demonstrate the substantial tunability of the magnetoelastic constant by nearly 50% via gate-induced strain. Our findings demonstrate the strong spin-mechanical coupling in CrSBr and paves the way for developing sensitive magnetic sensing and efficient quantum transduction at atomically thin limit.

Keywords: Nanomechanical resonators, spin-mechanical coupling, magnetostriction, 2D magnets

## Introduction and Background

Recent progress in layered magnetic materials represents an emerging research frontier to explore new types of spin orderings and harvest their collective excitations for technological advancements. For example, various types of magnetic ordering have been discovered such as 2D ferromagnets/antiferromagnets[1,2,3], noncollinear spin textures[4,5], quantum spin liquid[6,7] and magnetic topological insulators[8,9]. Among them, 2D A-type antiferromagnet CrSBr has garnered significant attention due to substantially enhanced spin excitations[10,11], strong couplings with excitons[11–14] and superior air stability[15,16]. A CrSBr crystal is composed of rectangular unit cells arranged in layers within the *ab*-plane, and these layers are sequentially stacked along the *c*-axis, resulting in an orthorhombic structure (Figure 1a). It is found to be A-type antiferromagnetic material below a Néel temperature $T_N$ around 132 K[17]. Spins within each layer align ferromagnetically along the crystalline *b* axis while the interlayer coupling is antiferromagnetic. Recent advance has reported strong magnon-exciton coupling, where the exciton energy can be nontrivially modified by the interlayer spin alignment by about 10 meV due to spin-dependent exchange interaction, enabling new efficient mechanism for magnon-exciton transduction[11]. Subsequent research has provided compelling evidence of the ultrastrong couplings between cavity photons and excitons within CrSBr[13,14], which is previously inaccessible in bulk materials. Moreover, unlike most 2D magnets, ultrathin CrSBr is air-stable and can last for months in ambient conditions[15,16], which is most appealing for practical device applications. These findings highlight that CrSBr can lead to unique magnetic phases[18] and highly tunable quantum information carriers[19] for low-power spintronics and hybrid magnonics.

Besides the coupling between charge and spin degree of freedoms, the interplay with lattice degree of freedom in quantum materials is also important for correlated physics and hybrid magnonics. One notable example is the magnetostriction effect, which dictates how lattice deformation accompanies with magnetization change[20–22]. The underlying spin-mechanical coupling is found to be crucial for magnon generation and transport[23,24], on-demand modulation of magnetism[25,26], efficient quantum transduction[27,28,22], and high-performance sensors and actuators[29,30]. However, such an important effect and its great potential in hybrid magnonics is largely unexplored in 2D CrSBr and other 2D layered magnets due to lack of sensitive mechanical probing methods at ultrathin limit and micrometer scale. For example, conventional high-sensitivity measurement techniques for magnetostriction such as strain gauge, capacitance dilatometry and cantilever measurements[31], are challenging to be applied to the microscale 2D samples due to the geometrical incompatibility and low signal-noise ratio.

Here we address this challenge and interrogate the magnetostriction and magnetoelastic coupling in ultrathin CrSBr membranes using nano opto-electro-mechanical resonators, whose high-quality mechanical and optical cavity allowing for sensitive mechanical detection. A typical nanomechanical system is a nanoscale device made by a thin membrane, whose nanomechanical vibration frequency and amplitude can be precisely probed by optical interferometry[32]. Its nanomechanical resonance response can contain critical thermodynamic and magnetoelastic properties of the material. Specifically, entropy changes in a material arise from magnetic order reorientation and transitions, are reflected in the modification of its specific heat. This modification leads to variations in the thermal expansion coefficient that affect the tension and resonance frequency[33,34]. Building upon the sensitive optical interferometry and high-quality mechanical cavity, we reveal the thermodynamic properties of the magnetic phase transition and interrogate the nontrivial magnetostriction effect in multilayer CrSBr. The distinct magnetostriction behaviors observed under different magnetic fields and temperatures constitute a B-T phase diagram of spin alignment and magnetic ordering. Based on a free energy model for magnetostriction, we quantify the saturation magnetostriction $\lambda_s$ to be $2.3 \times 10^{-5}$ and magnetoelastic coupling constant to be $3.4 \pm 0.1$ MJ/m$^3$. The saturation magnetostriction in CrSBr is one order of magnitude larger than that of yttrium iron garnet (YIG)[35], the state-of-the-art quantum material for hybrid quantum magnonics. Furthermore, we have successfully shown that the magnetostriction effect in multilayer CrSBr can be extensively controlled by gate-induced strain, with the magnetoelastic coupling strength exhibiting as high as 50% amplitude tuning. Our finding may unleash the full potential for CrSBr as new hybrid magnonic materials and pave the way for using lattice, spin, and charge degree of freedoms in it for quantum transduction.

## Integrated Results and Discussion

To fabricate the nano opto-electro-mechanical cavity systems (NOEMS), we exfoliated multilayer CrSBr flakes from synthesized bulk crystals (Supplementary Figure S1) to the Polydimethylsiloxane (PDMS) polymer and then transferred onto the pre-patterned circular hole (Figure 1c, see Method for more details). A DC gate voltage $V_g$ coupled through the bias tee can change the pre-tension in the membrane. On the other hand, the vibration of the membrane is excited by a small RF voltage generated by a vector network analyzer and detected interferometrically with a 633 nm He-Ne laser (Figure 1b). When the driving frequency provided by VNA matches the mechanical resonant frequency of the suspended membrane[32], a peak emerges in the microwave transmission spectrum (Figure 1d). The high-quality factor ($Q \approx 5000$) enables high sensitivity for the following mechanical detection of spin orderings and magneto-elastic coupling.

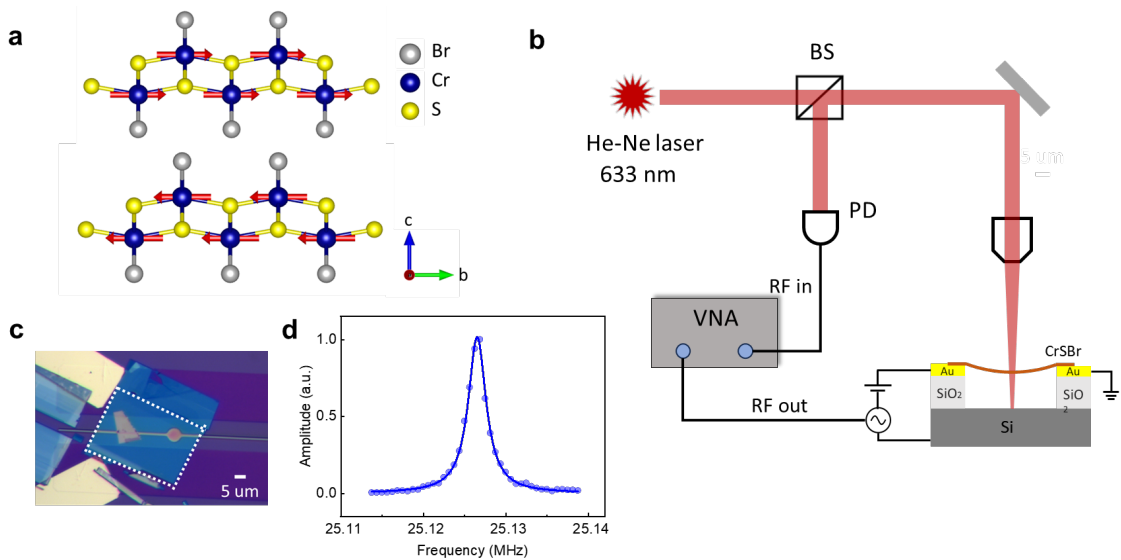

**Figure 1: Nano opto-electro-mechanical resonators with multilayer CrSBr.** (**a**) lattice structure and the spin alignment of CrSBr. (**b**) Schematic of the measurement system. The resonator is electrically actuated by a vector network analyzer (VNA). A DC gate voltage $V_g$ is superimposed to apply static tension to the membrane through a bias tee. The motion results in dynamic boundary change and optical interference of incident He-Ne laser beam, whose intensity modulation at different vibration frequency can be detected by an ultrafast PD and the same VNA. BS: beam splitter; PD: photodetector. (**c**) Optical microscope image of a CrSBr opto-electro-mechanical resonator. Part of multilayer CrSBr flake (white dashed line) is suspended over an etched circular drum as mechanical resonator, the Au/Ti electrodes are prepatterned to enable electrical contact with the flake. (**d**) Typical amplitude curve versus the driving frequency measured at 1.7 K, showing a resonance peak at around 25.126 MHz and a quality factor around 5000. The blue curve is the Lorentz fit of the data points.

To reveal the spin-mechanical couplings in different magnetic orderings, we first use NOEMS to detect the magnetic phase transition and corresponding thermodynamic properties in multi-layer CrSBr with a thickness of 30 nm (See Supplementary Figure S2.1). In particular, we measured the temperature dependent mechanical resonance frequency $f_0(T)$ in the range from 117 K to 145 K, (solid blue curve in Figure 2a). We observed a smooth frequency redshift except for a subtle abrupt change around 139 K. This signature is more evident by $df_0^2(T)/dT$ (solid red curve in Figure 2a), which shows a sudden large dip around 139 K. The gradual thermal expansion induced strain change is commonly accounted for the smooth frequency decrease, given the resonant frequency is highly dependent on membrane strain ($f =$

$\frac{2.4}{2\pi R}\sqrt{\frac{E_Y \varepsilon_r}{\rho}}$, where $E_Y$ is Young's modulus, $\varepsilon_r$ is strain, $R$ is the radius of the circular resonator and $\rho$ is the mass density). While the sudden frequency slope change around 139 K is attributed to the antiferromagnetic to paramagnetic transition in CrSBr. Here the magnetic entropy changes results in subtle tension variation, which is captured by the ultrahigh frequency sensitivity in our NOEMS approach. Note that the $T_N$ detected in our ultrathin CrSBr membrane is slightly higher than its bulk counterpart (132 K), which is assigned to a distinctive intermediate or surface magnetic phase[36,37]. In addition, the quality factor of resonance, defined as the ratio between resonance frequency and resonance linewidth, is a key quantity reflecting the mechanical loss in a mechanical resonator. In our case, it gradually decreases from 117 K with a minimum occurring near 139 K, which coincides with the Néel temperature obtained from the previous $f_0(T)$ curve (Figure 2b). It suggests that larger dissipation near the phase transition due to more significant thermoelastic damping[38,39]. The identical critical temperature again confirms the magnetic phase transition around 139 K in ultrathin CrSBr NOEMS device. Moreover, the heat capacity $c_v$ is proportional to the derivative of resonance frequency squared with respect to temperature ($c_v \propto df_0^2(T)/dT$) [33] (see more details in Supplementary Information Section S4). Accordingly, we plotted the value of heat capacity through the magnetic phase transition. The second order magnetic phase transition at Néel temperature features an anomaly in heat capacity. And our calculated values are consistent with those reported for bulk CrSBr crystals[16,37]. This demonstrates the feasibility of our NOEMS approach in measuring thermodynamical property in $\mu$m-scale 2D materials, which is challenging for conventional calorimetry method[40].

Building upon the magnetic phase diagram, we investigated the magnetostriction effect in different spin orderings of CrSBr. To start, the mechanical resonance $f_0$ of the CrSBr device was measured under an out-of-plane magnetic field scanning at 1.7 K (Figure 3a). We observed that $f_0$ redshifts smoothly with increasing field at 1.7 K and saturates beyond 2.08 T. The total change in $f_0$ is about 0.05 MHz or 0.29% ($\frac{f_{AFM}-f_{cAF}}{f_{AFM}} \approx 0.29\%$) of the initial frequency. Since the mechanical resonance variation is directly connected with strain change in the membrane, such nontrivial frequency shift under magnetic field is a hallmark of magnetostriction effect in the same ultrathin CrSBr device shown in Figure 2. Moreover, the observed saturation field (2.08 T) is consistent with the value reported for multilayer CrSBr[41,42] and our magnetic circular dichroism data (Supplementary Section S7 and S8). Thus, we attributed the smooth mechanical frequency redshift is correlated with the spin-canting from in-plane antiferromagnetism to out-of-plane ferromagnetism in 2D CrSBr. During the process, the spin alignment changes the strain in the membrane via magnetostriction effect.

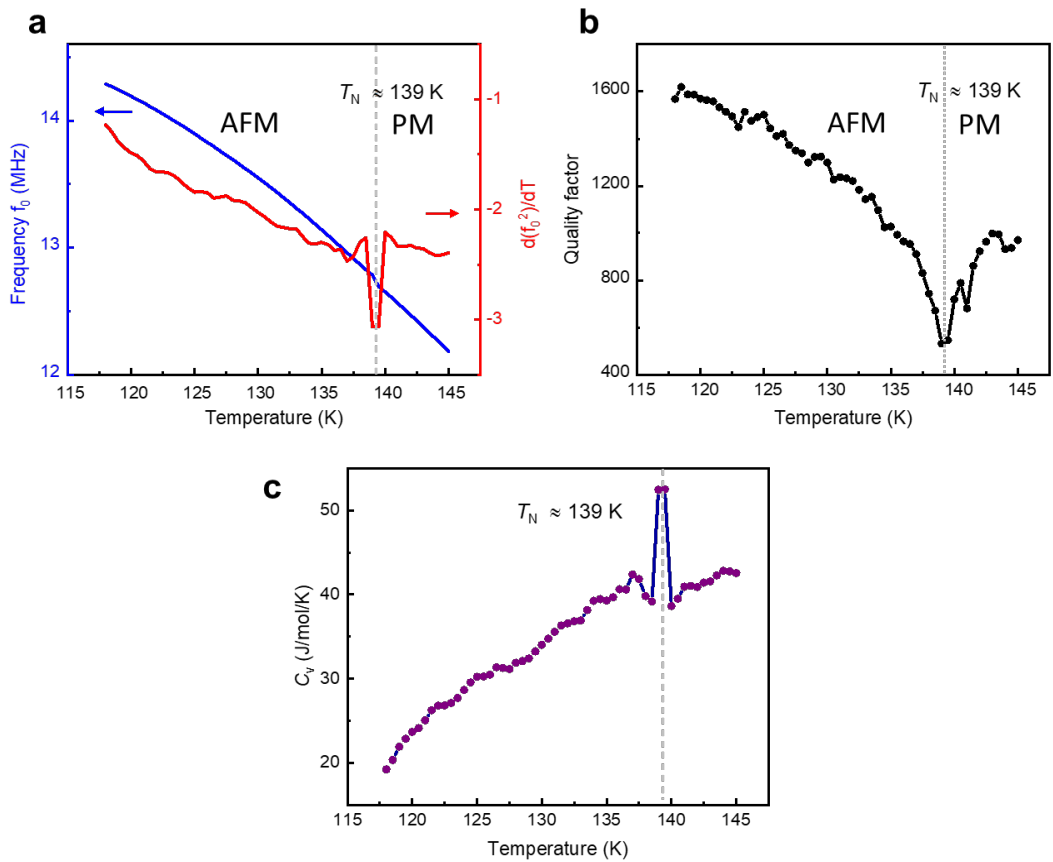

**Figure 2**: **Mechanical detection of magnetic phase transition and thermodynamic properties of multilayer CrSBr.** (**a**) Detection of magnetic phase transition temperature. Solid blue curve - resonance frequency as a function of temperature, solid red curve - temperature derivative of $f_0^2(T)$, with a minimum at 139 K, which corresponds to Néel temperature $T_N$. **(b)** Temperature dependence of mechanical quality factor $Q$ of the fundamental resonance mode, the drop of $Q$ factor at AFM-PM magnetic phase transition suggests the larger mechanical loss arises from spin fluctuation. **(c)** Heat capacity $c_v$ of CrSBr thin flake calculated from $f_0(T)$, exhibits an anomaly at 139 K due to magnetic phase transition.

To quantify the magnetoelastic coupling strength, we have adapted the conventional magnetostriction model[34,43], with the specific CrSBr magnetic and elastic free energy to fit our experimental data. Our model indicates that the competition between minimizing the internal magnetic energy and elastic energy contributes to the resonance frequency shifts with magnetic state. The elastic energy of the membrane per unit volume can be expressed as $U_{el} = \frac{3}{2} E_Y \varepsilon^2$, where $\varepsilon$ is the strain, $E_Y$ is Young's modulus. On the other hand, the intrinsic magnetic energy in CrSBr can be written as[44]:

$$H = \sum_{n-1} -J_\perp S_t \cdot S_b + \sum_n [-D(S_i^y)^2 + E[(S_i^z)^2 - (S_i^x)^2]]$$

Here $J_\perp$ is interlayer exchange coupling with energy per unit volume. $S_t$, $S_b$ denotes the spin unit vector of the top and bottom CrSBr layers, and the single-ion anisotropy parameters $D$ and $E$ are introduced to simulate the triaxial magnetic anisotropy[44]. The two terms of CrSBr spin Hamiltonian describe the contribution from the magnetic exchange energy and anisotropic energy, respectively. By minimizing the total energy with respect to strain, we can obtain the expression of frequency shift as a function of the applied magnetic field. (See Supplementary Information Section S5 for more details)

$$f_{AFM}^2 - f_{cAF}^2 = \frac{0.82}{\rho_{eff} \pi^3 R^2} B_{me} \sin^2\theta$$

Here the magnetoelastic constant $B_{me} = (n-1)/n \cdot \partial J_\perp/\partial \varepsilon - 1/2 \cdot \partial(E+D)/\partial \varepsilon$, $n$ is the layer number of CrSBr devices. When $n \gg 1$, $B_{me} \approx \partial J_\perp/\partial \varepsilon - 1/2 \cdot \partial(E+D)/\partial \varepsilon$, $\rho_{eff}$ is the effective mass density, $\sin\theta = M/M_s = H/H_s$ for the antiferromagnetic phase. $H_s$ is the magnetic saturation field. From this model we found the frequency shift within spin canting process can be well fit (Supplementary Information Figure S3), and we can further infer the magnetoelastic constant $B_{me} = 1.8 \pm 0.1$ MJ/m³, here the value derivation is the fitting error bar. By calculating the maximum strain change the under magnetic field based on the resonance frequency, we also determined a saturation magnetostriction coefficient $\lambda_s = 1.3 \times 10^{-5}$ at 1.7 K.

To fully understand the magnetostriction in different spin orderings, we have conducted a systematic temperature dependent magnetostriction study (Figure 3b). As the temperature rises, the magnetic saturation field drops arising from a weakened magnetic exchange interaction. Accordingly, the temperature dependence of magnetoelastic constants in antiferromagnetic regime declines dramatically as the temperature increases from 1.7 K to 120 K (Figure 3c). Besides, resonance frequency blueshift beyond the saturation field, especially at higher temperature in contrast to a nearly flat curve observed at 1.7 K. Above the Néel temperature, the magnetic saturation field is no longer observable, and $f_0$ blueshifts continuously with an increasing magnetic field (Figure 3c, 170 K and 200 K data). The large magnetostriction in paramagnetic phase is attributed to the large magnetocrystalline anisotropy generated by the spin orbit coupling (SOC) of the bromine atoms[41], similar to what observed in other magnetostriction materials[45]. The different nature of magnetostriction effect in CrSBr at different regimes enables us to draw a magnetic phase diagram with respect to temperature and applied magnetic field (Figure 3d). The intensity of the mapping plot is indicated by the magnetostriction sensitivity defined as $\partial\varepsilon/\partial B$, which characterize the magnetostriction strength[46]. The transition between the FM and AFM state is distinguishable due to the discontinuity of $\partial\varepsilon/\partial B$ across the two magnetic phases. Conversely, in the PM state, the strain increases monotonically with the magnetic field and no discontinuity is observed.

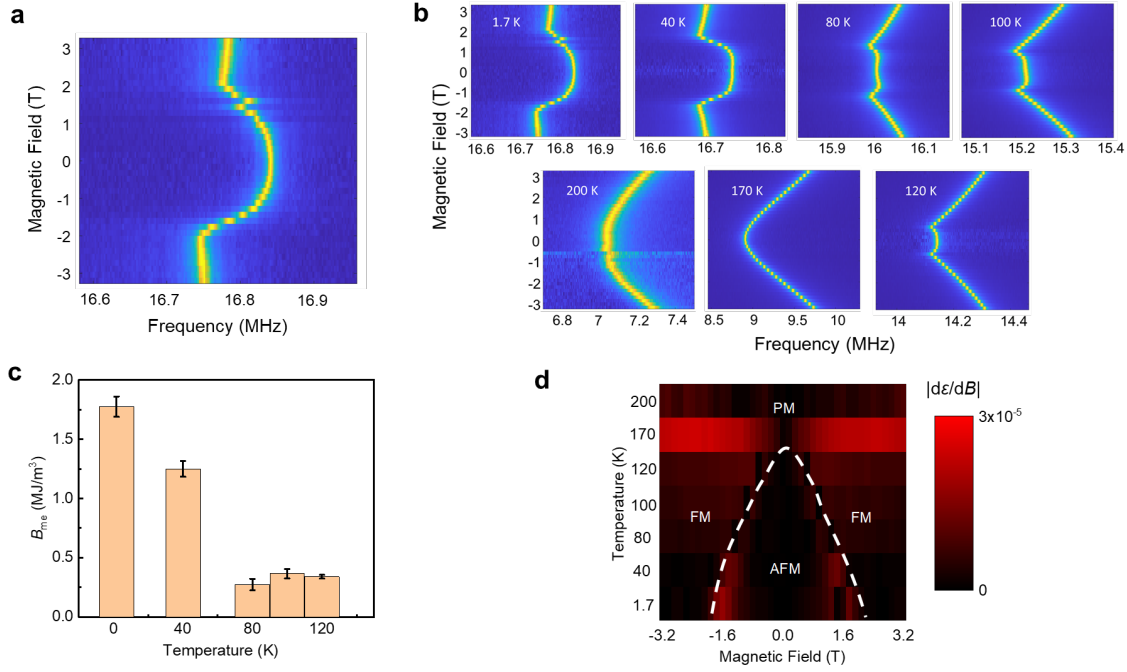

**Figure 3: Probing magnetostriction effect in different magnetic regimes.** (**a**) Normalized vibration amplitude as a function of driving frequency and out-of-plane magnetic field that sweeps from -3.2 T to 3.2 T, measured at 1.7 K. The resonance frequency blueshifts as magnetic field increases until the magnetization is saturated. The magnetic field dependent mechanical resonance frequency shift clearly reveals the nontrivial spin-mechanical coupling. (**b**) Temperature dependent magnetostriction effect at 1.7 K, 40 K, 80 K, 100 K, 120 K, 170 K, 200 K, respectively. (**c**) The magnetoelastic constant in antiferromagnetic ordering as a function of temperature fitted from the magnetostriction model, which drops dramatically as the temperature increases from 1.7 K to 120 K (**d**) Magnetic phase diagram of CrSBr using the absolute value of $\partial\epsilon/\partial B$ as an indicator. The dash white line delineates the boundary between antiferromagnetic (AFM), ferromagnetic (FM) and paramagnetic (PM) states.

Finally, we interrogated the tunability of magnetostriction effect in another multilayer CrSBr (15 nm, Supplementary Figure S2.2) NOEMS device by applying in-situ DC gate bias between suspended membrane and silicon substrate. In particular, we measured the magnetic field dependent resonance frequency at different DC gate voltages (Figure 4a and Supplementary Figure S7). As the gate voltage increases, we observed the significant drop of frequency change between zero field and saturation field in the $f-H$ curves, which suggests a tunable magnetostriction effect. Through the parabolic fitting of the $f-H$ curve using established magnetostriction model, we quantified magnetoelastic constant $B_{\mathrm{me}}$ as a function of gate voltages (Figure 4b). The $B_{me}$ for this device at zero gate voltage is $3.4 \pm 0.1$ MJ/m$^3$, and magnetostriction coefficient $\lambda_s = 2.3 \times 10^{-5}$. This value is one order of magnitude larger than yttrium iron garnet (YIG), and comparable to iron[35,47]. We notice that the $B_{me}$ here is larger than that in the 30 nm device. In several other devices we measured, the fitted $B_{me}$ range from 2 – 4 MJ/m$^3$ (See Supplementary Table 1), and no clear thickness dependence is observed. As a result, we suspect multiple extrinsic factors during device fabrication, including initial transfer strain and polymer residue on the CrSBr membrane, may lead to this variation. This observation of large magnetoelastic coupling is supported by our first-principles density functional theory (DFT), which suggests similarly large value in 2D vdW CrSBr (Supplementary Section S10). Furthermore, we found that $B_{me}$ decreases remarkably as the voltages increases to 30 V with up to 50% amplitude tunability. The tuning trend is almost symmetrical for positive and negative voltages.

In the following, we will discuss the possible underlying tuning mechanisms. On one hand, the DC gate bias can provide additional capacitive force to pull the suspended CrSBr membrane downwards and build up extra tensile strain to alter magnetostriction. On the other hand, the DC gate bias can also induce carrier doping into the membrane, which may modify the exchange and the anisotropy energy terms in the magnetostriction model. To elucidate the observed tunability contributed by these two mechanisms, we have conducted both reflective magnetic circular dichroism (RMCD) and DFT calculations. Firstly, we implement RMCD on the supported region of the same CrSBr flake on the SiO$_2$/Si substrate to quantify the gate-dependent saturation field (Supplementary Figure S4), in which the gate bias can only induce electrostatic doping. We noticed that the saturation field $B_s$ = 20400 Oe remains unchanged for CrSBr flake on the substrate under gate voltages up to 90 V. Note that $B_s$ is linearly proportional to interlayer exchange interaction $J_\perp$ and magnetic anisotropy energy, and its first derivation respect to strain determines the magnetoelastic constant $B_{me}$. Therefore, the observed nearly unchanged $B_s$ for ultrathin supported CrSBr suggests that the doping effect is negligible to alter the magnetoelastic coupling by up to 50%. Our DFT calculation results also point out the gate induced doping (~ $10^{11}/cm^2$), has a minor effect on the interlayer exchange interaction and magnetic anisotropy energy (Supplementary Information Section S10).

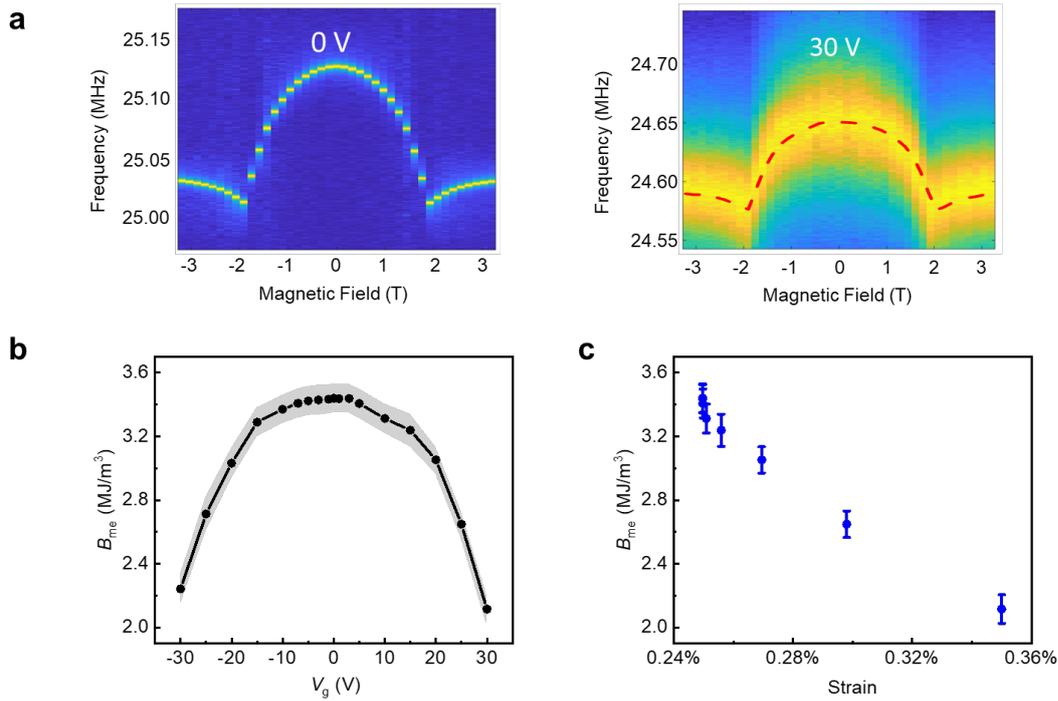

**Figure 4: Gate Tuning of the magnetostriction effect in CrSBr membrane.** (a) Resonance frequency with respect to magnetic field under gate voltages 0 V and 30 V measured at 1.7 K. Frequency change between zero field and saturation field at 30 V is significantly smaller than that at 0 V, indicating a tunable magnetoelastic coupling. (b) Fitted magnetoelastic coupling constants $B_{me}$ against gate voltages from -30 V to 30 V. $B_{me}$ decreases symmetrically with increasing $|V_g|$, achieving up to 50% strength change. The shaded region represents the error band from the fitting. (c) Fitted magnetoelastic coupling constants $B_{me}$ as a function of strain. Given the nearly symmetric response, only the data of strains induced by positive voltages are plotted for simplification.

Regarding the gate-induced strain influence, we observed $B_s$ decreases from 19000 Oe to 18400 Oe within the same gate voltage range by RMCD measured on the suspended drum

(Supplementary Figure S5). Based on this $B_s$ difference, we calibrate the strain change according to the strain dependent saturation field[48], and found a gate-induced tensile strain on the order of 0.1% under the 30 V gate voltage (Figure 4c and Supplementary Figure S6). The observed strain tunability of magnetoelastic coupling is higher than conventional bulk magnetostriction materials such as Fe, Ni and Co[49,50]. For example, in epitaxial Fe(001) films, around 0.5% strain is required to tune $B_{me}$ by 50% [49], while only 0.1% strain in needed for CrSBr to achieve same relative magnitude change. Such a large strain tunability of magnetoelastic coupling and magnetostriction may be understood by the strong modification of geometry of the Cr-Br-Br-Cr exchange pathway and the distinctive mechanical flexibility of 2D membranes, where strain tuning of electronic orbital change results in substantial spin responses. Besides, a large shape anisotropy arising from the in-plane spin orientation may also play a role here[41]. Taken together, our finding, for the first time features the discovery of substantial tunability of spin-mechanical coupling in 2D magnetic membranes.

## Concluding Remarks

Our results of magnetostriction effect in ultrathin CrSBr suggest its great potential as superior quantum magnonics platform for information transduction and spintronics. For example, compared to YIG, the state-of-the-art material for hybrid quantum magnonics, the demonstrated three-times larger magnetostriction and much stronger magneto-optic effect in atomically-thin CrSBr can potentially overcome the weak quasiparticle coupling and the large material mode volume challenges, leading to long-sought efficient quantum transduction using cavity magnomechanics and cavity optomagnonics schemes[51,52]. Such nontrivial magnetoelastic coupling in CrSBr can also lead to efficient magnon generation at 2D limit via planar surface acoustic wave launching[53,54], which is important for on-chip spintronics. Moreover, the highly tunable magnetoelastic coupling strength in CrSBr is beneficial when exploring quantum critical phenomena across different quantum phases. For instance, the observation of exceptional points and exceptional surface in non-Hamiltonian magnon polariton systems benefits a large tunable coupling strength to achieve the degeneracy of both eigenfrequencies and eigenvectors. The large turnability also allows for illustrating the evolution of Riemann surfaces associated with real and imaginary parts of the eigenvalues as the coupling strength changes[55,56].

In summary, we have studied the magnetic phase transition and magnetostriction effect in 2D layered magnets CrSBr using high-quality nano opto-electro-mechanical resonators. The magnetic phase transition around $T_N$ and associated thermodynamical property such as specific heat was characterized through the temperature-dependent resonance frequency analysis. Through magnetic field dependent mechanical frequency measurements and magnetostriction model fitting, we found distinct magnetostriction effect $\lambda_s = 2.3 \times 10^{-5}$ and magnetoelastic coupling strength on the order of $10^6$ J/m$^3$. Furthermore, we have demonstrated significant gate-induced strain tunability ~ 50% of magnetoelastic coupling. A model for magnetostriction was applied to quantify the magnetoelastic constant $B_{me}$, whose strain tunability in CrSBr is found to be much larger than the typical thin film magnetic materials such as cobalt and iron. Distinct from prior magnetostriction studies in 2D magnets CrI$_3$ and MnBi$_2$Te$_4$[34,57], our results on CrSBr have unique significance in terms of: (1) distinctive material platform with strong coupling physics. Specifically, CrSBr features a multitude of quasiparticle interactions including magnon-exciton, magnon-phonon, magnon-magnon and cavity photon-exciton couplings; (2) superior air stability. In contrast to extremely air-sensitive CrI$_3$ and MnBi$_2$Te$_4$, atomically-thin CrSBr flakes are robust under the ambient condition[15], which is critical for routine and reliable device applications. (3) CrSBr also has a much higher magnetic transition temperature[16] above the liquid nitrogen temperature with lower cryogenic operation cost. Taken together, our study advances the understanding of spin-mechanical coupling in 2D layered quantum materials and paves the way for low-dimensional magnetostriction device applications such as spintronics, magnetic sensing and quantum transduction.

## Acknowledgement

This research was primarily supported by NSF through the University of Wisconsin Materials Research Science and Engineering Center (DMR-2309000). The work at University of Texas at Dallas is supported by the US Air Force Office of Scientific Research (AFOSR) (FA9550-19-1-0037), National Science Foundation (NSF) (DMREF-2324033) and Office of Naval Research (ONR) (N00014-23-1-2020).

## Author Contributions

J.X. supervised the project; J.X. and F.F. conceived the research and designed the experiments; F.F. performed the laser interferometry measurements and analyzed the data with J.X.; Y. M. fabricated the devices with assistance from F.F. and J.R., under the guidance of Y.W. and J.X.; W.F. conducted first-principles calculations supervised by Y.P.; W.L. and A. K. synthesized the bulk CrSBr crystal under the guidance of B.L.; All authors discussed the results and jointly wrote the paper with a main contribution from F.F. and J.X.

## Competing interests

The authors declare no competing interests.